\documentclass{mem}
\usepackage{natbib}\usepackage{txfonts}\usepackage{balance}
\usepackage{graphicx}
\usepackage[a4paper]{hyperref}
\idline{75}{282}
\begin{document}
\def\teff{$T\rm_{eff }$}
\def\kms{$\mathrm {km s}^{-1}$}

\title{Are there cool-core clusters at high-redshift? \\
 \textit{Chandra} results and prospects with \textit{WFXT}}

\author{
J.S. \,Santos\inst{1}, P. \,Tozzi\inst{1}
\and P. \, Rosati\inst{2}
          }

\institute{
INAF--Osservatorio Astronomico di Trieste, Via Tiepolo 11,
I-34131 Trieste, Italy \\
 \email{jsantos@oats.inaf.it}
\and
European Southern Observatory}

\authorrunning{Santos et al.}

\titlerunning{Evolution of cool-cores: \textit{Chandra} results and prospects with \textit{WFXT}}

\abstract{
Cool core clusters are characterized by strong surface brightness peaks in the X-ray 
 emission from the Intra Cluster Medium (ICM). This phenomenon is associated with 
complex physics in the ICM and has been a subject of intense debate and 
investigation in recent years. 
The observational challenge of analyzing high redshift clusters and the small sample 
statistics have prevented an accurate assessment of the population of cool-cores at $z>$0.5.
In this contribution we trace the evolution of cool-core clusters out to $z\sim1.3$ using high-resolution \textit{Chandra} 
data of three representative cluster samples spanning different redshift ranges. 
Our analysis is based on the measurement of the surface brightness
(SB) concentration, $c_{SB}$, which strongly anti-correlates with the
central cooling time and allows us to characterize the cool-core
strength in low S/N data.  We confirm a negative evolution in the
fraction of cool-core clusters with redshift, in particular for very
strong cool-cores.  Still, we find evidence for a large population of
well formed cool-cores at $z \sim 1$.  This analysis
is potentially very effective in constraining the nature and the
evolution of the cool-cores, once large samples of high-z clusters
will be available.  In this respect, we explore the potential of the
proposed mission Wide Field X-ray Telescope (WFXT) to address this
science case.  We conclude that WFXT provides the best trade-off of
angular resolution, sensitivity and covered solid angle in order to
discover and fully characterize the cool-core cluster population up to
$z$=1.5.

}
\maketitle{}

\section{Introduction}

The majority of local X-ray clusters show a prominent central surface
brightness peak in the intra cluster medium (ICM). The cluster core is
also associated with a short cooling time, implying the presence of a
cooling flow (\citealt{fabian94}), although the gas is not observed to
cool below a minimum temperature of the order of $1/3$ of the average
value in the ICM, indicating that some heating mechanism counteracts
the cooling process.  The properties and the formation mechanism of
these cool-cores (CC) are an open problem which forces one to consider
complex non-gravitational physical processes able to provide smoothly
distributed heating on scales of about 100 kpc.  A successful model is
expected to include phenomena such as removal of radiatively cooled gas, heating by a
central radio source, thermal conduction or other forms of feedback
(see \citealt{peterson05}).

The impact of cool-cores on the local cluster population has been
extensively studied for over a decade (\citealt{peres}).
X-ray observations have established that cool-cores dominate the local
clusters, with an abundance of 50 to 70\%, depending on the 
adopted definition of cool-core (e.g. \citealt{chen}, \citealt{dunn}, 
\citealt{hudson}).   

The evolution of cool-cores has been measured only up to redshift 0.4. 
\cite{bauer} reported that the fraction of cool-cores does not 
significantly evolve up to $z\thicksim~0.4$, since clusters in this redshift range have 
the same temperature decrement (about one-third), as the nearby CC's, and their central 
cooling times are similar.
The study of cool-cores at redshift greater than 0.5 is plagued by low 
statistics, and, so far, is limited to two works. Using the 400 Square Degree Survey 
(hereafter 400 SD, \citealt{burenin}) which reaches $z=0.9$, \cite{vikhlinin06} concluded, 
on the basis of a cuspiness parameter defined as the logarithmic derivative of the 
density profile, that there is a lack of cool-core clusters, with respect to the local 
cluster population. 
In \cite{joana}, we adopted a simple diagnostic based on the concentration of the surface brightness 
(which strongly anti-correlates with the central cooling time), and 
measured the fraction of cool-cores out to the current redshift limit ($z \sim 1.4$).  
At variance with previous results, we found a significant 
fraction of what we term moderate cool-cores. 

In this contribution we present our results on the abundance of cool-cores across the 
entire cluster population, out $z\sim 1.3$, exploiting all the available data in the 
\textit{Chandra} archive (\citealt{joana10}), 
and we assess the potential of the next-generation X-ray mission Wide 
Field X-ray Telescope (WFXT, \citealt{giacconi}) in measuring the cool-core evolution.

\section{Cluster samples} 

The local cluster sample used in this work is drawn form the
catalog of the 400 Square Degree (SD) Survey (\citealt{burenin}), an 
X-ray survey which detected 266 confirmed galaxy clusters, groups or 
individual elliptical galaxies out to $z \sim 1$ using archival ROSAT 
PSPC observations. The sample is complete down to a flux limit of 
$1.4 \times 10^{-13}$ erg s$^{-1}$ cm$^{-2}$.
We extract a subsample of 26 clusters observed with {\sl Chandra} with $z>$0.05, 
in order to be able to sample the surface brightness profiles out to a radius of 
400 kpc within the field of view. Hence, our local sample spans the redshift
range [0.05 - 0.217].

X-ray images of distant clusters suffer a strong surface brightness dimming 
($\propto (1+z)^{-4}$) and have a small angular size, thus the study of their 
central regions requires the sub-arcsecond resolution provided only by \textit{Chandra}.
Beyond redshift 0.5 there are only three X-ray complete cluster samples, all 
selected from ROSAT PSPC pointed observations.  
They  are: (i) the 400 SD high-$z$ sample which includes all 
clusters (20) from the 400 SD catalog with $z\ge 0.5$; (ii) the Rosat Deep Cluster 
Survey (RDCS, \citealt{rosati98}; and (iii) the Wide Angle ROSAT Pointed Survey (WARPS, 
\citealt{jones}).    
While the distant 400 SD sample has been fully observed with {\sl Chandra}, the RDCS and WARPS 
samples have been only partially observed with a {\sl Chandra} follow up.  For this 
reason, we merge them into the RDCS+WARPS sample, containing a total of 15 clusters.  

\section{Surface brightness concentration}

The simplest observational signature of the presence of a cool-core is
a central spike in the surface brightness profile.  This is also the
only possible diagnostic we can apply to high redshift clusters, given
the difficulty in performing spectral analysis to detect the
temperature decrease in the core region.  

In \cite{joana} we defined the phenomenological parameter
$c_{SB}$ that quantifies the excess emission in a cluster core by
measuring the ratio of the surface brightness within a radius of 40
kpc with respect to the SB within a radius of 400 kpc: $c_{SB} =
SB (r<40 kpc) / SB (r<400 kpc)$.  This simple parameter has been shown
to be robust and particularly useful when dealing with the low S/N data of 
distant clusters. We validated the redshift independence of $c_{SB}$
(apart from possible K-corrections as described in \citealt{joana10}) 
by cloning low-$z$ clusters to high redshift. After this detailed
investigation, we propose the use of $c_{SB}$ as the best proxy for the
cool-core stength in the high-z range.

\begin{figure}[]
\resizebox{\hsize}{!}{\includegraphics[clip=true]{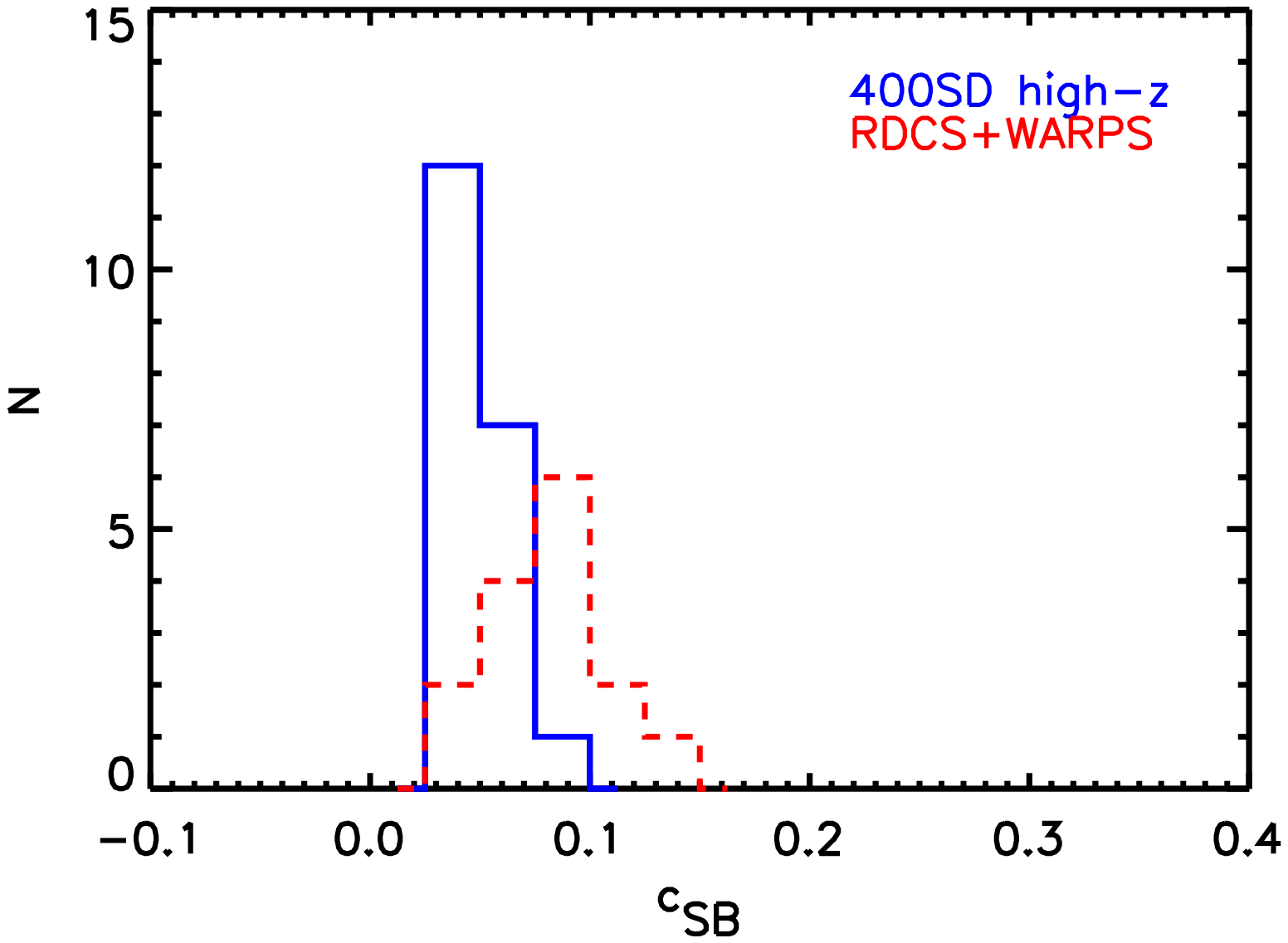}}
\resizebox{\hsize}{!}{\includegraphics[clip=true]{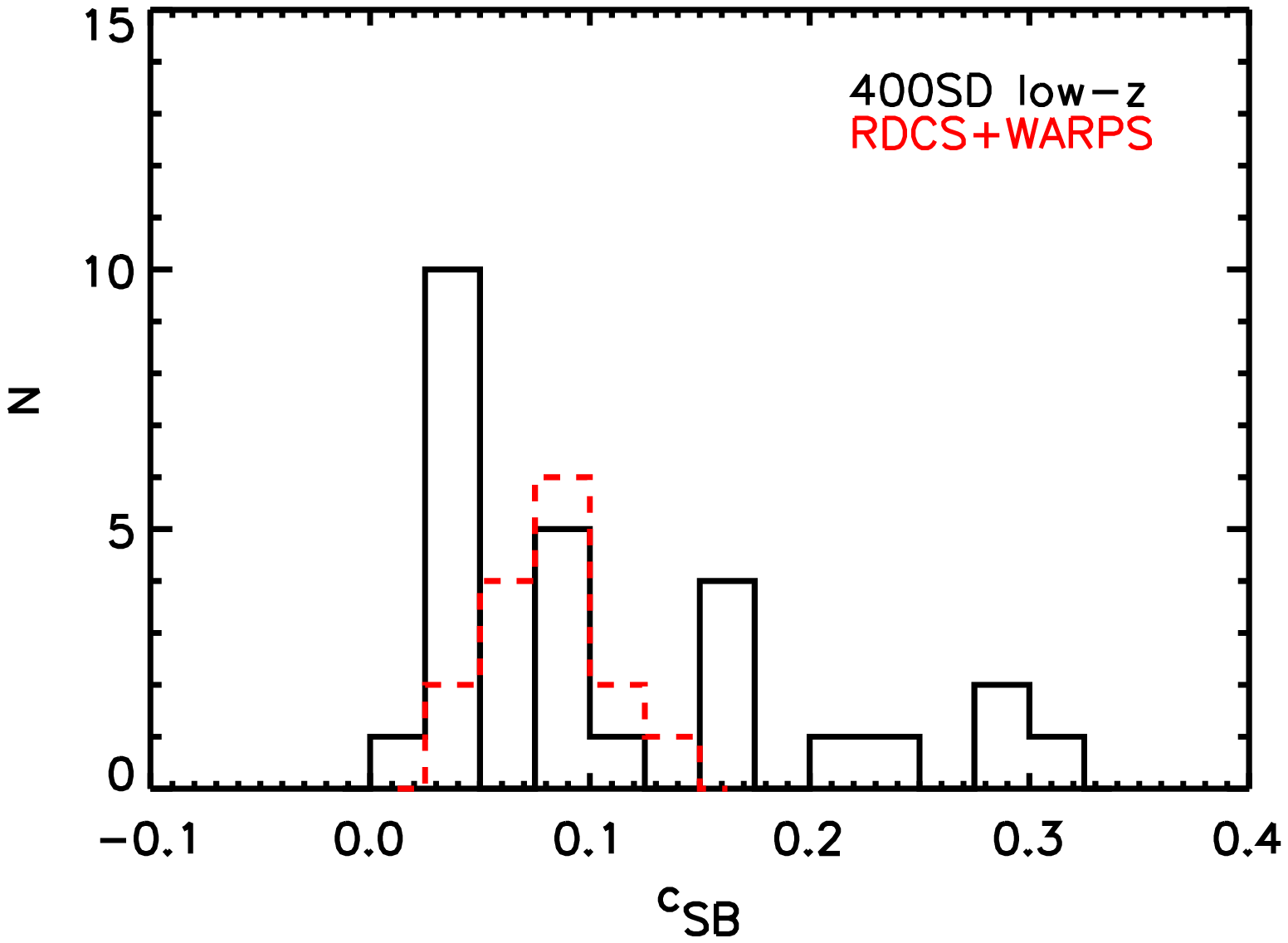}}
\caption{
\footnotesize
Comparison of the distribution of c$_{SB}$ of the distant samples (top) and of 
the local and RDCS+WARPS samples (bottom).}
\label{csb}
\end{figure}

Before comparing the $c_{SB}$ distribution of local and distant
samples (400 SD high-$z$ and RDCS+WARPS), we compare the two distant 
samples separately in Figure \ref{csb}, top panel.
Quite unexpectedly, the shape and range of the two high-$z$ $c_{SB}$
distributions are statistically different.  
We perform a K-S test and find a null hypothesis probability of
$0.6$\%, implying that the two distant samples do have different
distributions of cool-core strength.  The 400 SD high-$z$ reaches
$c_{SB}$ = 0.10, with median $c_{SB}=0.043$, whereas the RDCS+WARPS
reaches $c_{SB}=0.15$, with a median $c_{SB}$ value equal to 0.082.
The RDCS+WARPS clusters have thus a significantly higher surface
brightness concentration with respect to the 400 SD clusters. 

Since both the RDCS+WARPS and the 400 SD are samples based on ROSAT
data, we argue that this $c_{SB}$ difference is likely due to
different selection criteria used in the 400SD survey, resulting in a
bias against compact clusters with a relatively high  surface
brightness.
To check for these effects, we need to go through a detailed
comparison of the selection criteria in the three surveys, a task that
goes beyond the scope of this work.  In order to investigate the evolution
of the cool-cluster population, we decide to use the RDCS+WARPS only.

The $c_{SB}$ distribution of the local sample (Figure \ref{csb},
bottom) spans a broad range of values and reaches $c_{SB}$=0.315, with
a significant peak at low $c_{SB}$ and a median $c_{SB}$ equal to
0.079.  We performed a K-S test to the local and distant RDCS+WARPS
samples and found a null hypothesis probability of $16$\%, implying
that the two samples have a non-negligible probability to be
statistically similar. 
Our findings are compatible with a significant population
of cool-core clusters already well established at redshift $z\sim 1.3$
(5 Gyr after the Big Bang), while strong cool-cores ($c_{SB} > 0.150$)
must wait for a longer time span before they can develop.  To
reinforce these results, it is necessary to use larger samples of
high-$z$ clusters.

\section{Central cooling time}

The central cooling time is the measure most often used to quantify
cool-cores, as it provides a time-frame for the evolutionary state of
the gas.  Adopting an isobaric cooling model for the central gas,
$t_{cool}$ can be computed as:

\begin{equation}
t_{cool} = \frac{2.5n_{g}T}{n_{e}^{2} \Lambda(T) }   ,
\end{equation}
\label{tcool}

\noindent where $\Lambda(T)$, $n_{g}$, $n_{e}$ and T are the cooling
function, number density of ions and electrons, electron number
density and temperature, respectively (Peterson \& Fabian 2006).
Using the global cluster temperature we obtained the central cooling
time measured at a radius of 20 kpc.  The local clusters span a wide
range of ages [0.7 - 32.6] Gyr, whereas the RDCS+WARPS sample is
limited to [4.7 - 14.3] Gyr.

The local clusters span the wide range of ages [0.7 - 32.6] Gyr,
whereas the RDCS+WARPS sample is limited to [4.7 - 14.3] Gyr. The
fraction of clusters with a central cooling time lower than the age of
the Universe at the cluster redshift is: 58\% in the local sample;
27\% in the RDCS+WARPS, and 10\% in the 400 SD high-z.  However, a
more meaningful quantity would be the cooling time normalized to the
age of the cluster, defined as the time elapsed since the last major
merger event. Considering the age of the Universe at $z_{obs}$ is
misleading, as this is a loose upper bound on the age of the cluster.

We confirm a strong anti-correlation between $t_{cool}$ and 
$c_{SB}$ (see Fig.~\ref{tcool}), quantified by a Spearman rank test with 
coefficient $\rho$=-0.84.

\begin{figure}[]
\resizebox{\hsize}{!}{\includegraphics[clip=true]{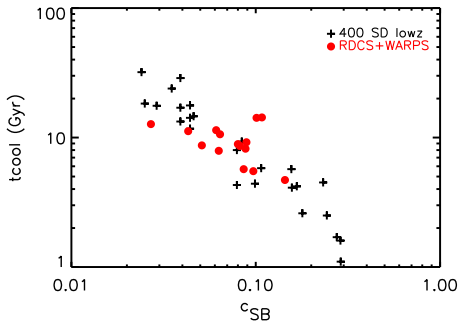}}
\caption{
\footnotesize
Correlation between central cooling time and the phenomenological parameter $c_{SB}$ for the 
local (crosses) and the high-z (filled cirles) samples.}
\label{tcool}
\end{figure}

\section{The potential of WFXT to measure the evolution of cool-cores}

With the present work we show that we can explore the population of
cool-core clusters up to the highest redshift where X-ray clusters are
selected, by exploiting the archive of \textit{Chandra}.  
This is possible thanks to the exquisite angular resolution of 
\textit{Chandra}, which allows us to sample the cool core region at any
redshift with about 10 resolution elements.  The only way to
improve the present work is to add serendipitously discovered high-$z$
 clusters followed-up with deep \textit{Chandra} observations. 
The number of $z>1$ X-ray clusters is slowly increasing as a result 
of the ongoing surveys with \textit{Chandra} and XMM-Newton.  However, sample 
statistics is not expected to increase significantly without a
dedicated wide area, deep X-ray survey.  Hence, it is instructive to
look into the future X-ray missions to investigate the capability of
characterizing the cool-core strength of high-redshift clusters.
Unfortunately, no proposed or planned future X-ray facility foresees an
angular resolution comparable to that of \textit{Chandra}. However, two future 
X-ray missions propose a PSF with a 5 arcsec half energy width (HEW) at 1 keV, 
the International X--ray Observatory and the Wide Field X--ray Telescope.

The International X-ray Observatory (IXO) (see e.g.,
\citealt{bookbinder}) is designed to have a great collecting power and
high spectral resolution, therefore it will provide very detailed
analysis of known or serendipitously discovered clusters, up to
high-redshift.  However, IXO will not be used in survey mode, but for
a limited solid angle, and therefore it would not increase
significantly the statistics of high-z cluster samples.

\begin{table}
\caption{\footnotesize Expected number of clusters sources with temperature $>$3 keV and minimum 
net counts 1500 in each of the three planned WFXT surveys in two redshift bins. }
\label{abun}
\begin{center}
\begin{tabular}{lll}
\hline
\textbf{Survey} & \textbf{0.5$<$z$<$1.0}  &  \textbf{ 1.0$<$z$<$1.5} \\
\hline
Shallow  & 200   & 0 \\
Medium   & 2190  & 300 \\
Deep     & 188   & 94 \\
\hline
\end{tabular}
\end{center}
\end{table}

\begin{figure*}
\resizebox{\hsize}{!}{\includegraphics[clip=true]{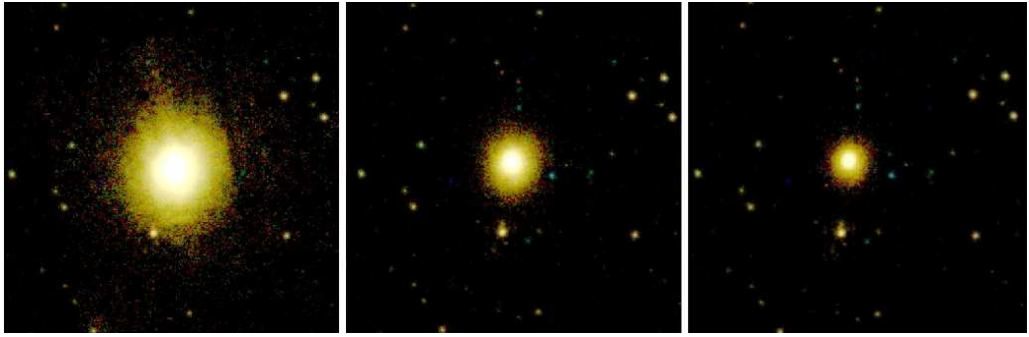}}
\caption{ \footnotesize Simulated WFXT images of the strong cool-core
  cluster A1835 in the medium survey (13.2 ksec), at redshifts 0.5
  (left), 1.0 (middle) and 1.5 (right).  The images are 10 arcmin
  across and are displayed in logarithmic scale.  }
\label{wfxt}
\end{figure*}

\begin{figure*}
{\includegraphics[clip=true,width=6.8cm]{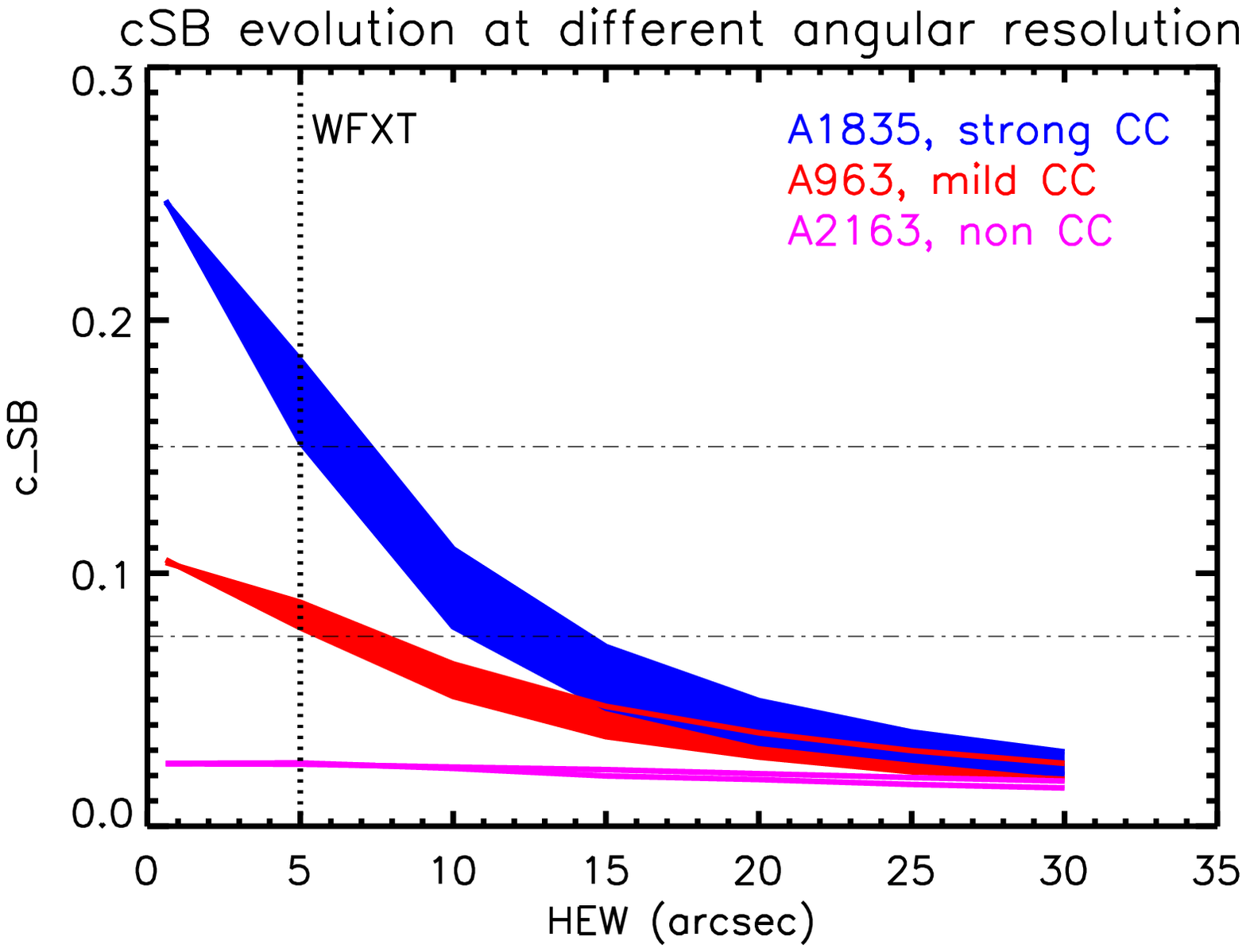}}
{\includegraphics[clip=true,width=7.1cm]{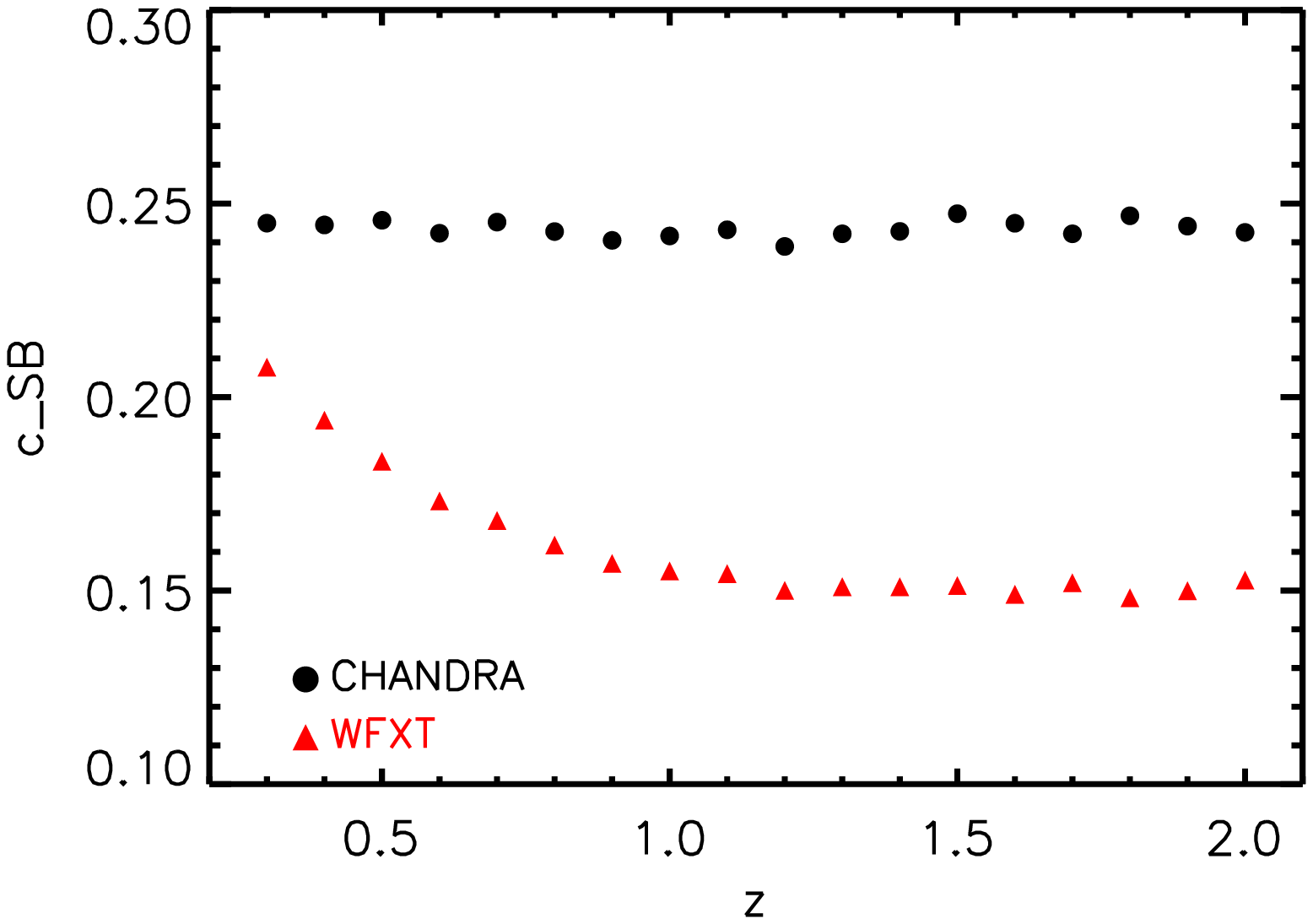}}
\caption{
\footnotesize
(Left) Variation of the measured $c_{SB}$ as a function of the telescope HEW for the three typical 
cases (strong-, moderate- and non-cool core).
For each cluster the coloured area is bounded by the $c_{SB}$ value at redshift 0.5 (higher bound) and 1.5 
(lower bound). The two horizontal dash-dot lines represent the boundaries between strong CC and 
moderate CC (upper line, $c_{SB}$=0.15) and moderate-CC and non-CC (lower line, $c_{SB}$=0.075).
(Right) Comparison between Chandra and WFXT measures of $c_{SB}$ as a function of redshift, for 
a strong cool-core cluster.
}
\label{wfxt2}
\end{figure*}

The Wide Field X-ray Telescope (WFXT) is one of the most promising
proposed X-ray missions.  The expected number of high-z clusters
detected in WFXT surveys with signal to noise comparable to that of
the cluster sample used in this work (conservatively expressed as a
lower bound of 1500 net counts), is shown in Table ~\ref{abun}.
Simulations of realistic WFXT fields have been produced in order to
investigate the accuracy of WFXT in characterizing cool core clusters.
We used the cloning technique (\citealt{joana}) to simulate WFXT
images of three canonical cluster types, corresponding to a typical
strong-CC ($c_{SB}>0.150$, A1835), a moderate-CC ($c_{SB}>0.075$,
A963) and a non-CC ($c_{SB}<0.075$, A2163) (see \citealt{joana} for
more details on these clusters), at redshifts 0.5, 1.0 and 1.5. To
obtain a quantitative assessment of the cool-core properties of the
simulated clusters, we measured $c_{SB}$ in the simulated images.

We first investigated the effect of the angular resolution in the
evaluation of the cool-core strength.  The result clearly shows that
the ability of an instrument to resolve the core and discriminate
between a cool-core and non cool-core is compromised for HEW greater
than 10\arcsec (see Figure ~\ref{wfxt2}, left panel).  In more detail,
the $c_{SB}$ values for the strong cool-core cluster A1835
(Fig.~\ref{wfxt}) as measured by WFXT are shown in the right panel of
Figure ~\ref{wfxt2}, in comparison with the values measured by
\textit{Chandra}.  We notice an apparent evolution in $c_{SB}$ due to
the larger angular resolution of WFXT relative to \textit{Chandra},
but we also confirm that the measure of $c_{SB}$ at face-value allows
us to assign the different clusters to their own cool-core class
(i.e. strong, moderate or non cool-core) at any redshift, as already
shown in Figure~\ref{wfxt2}.
\begin{figure*}
\resizebox{\hsize}{!}{\includegraphics[clip=true]{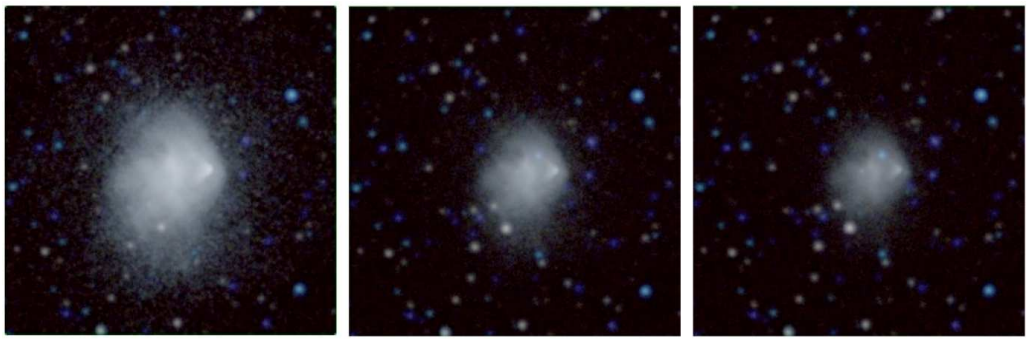}}
\caption{ \footnotesize The Bullet cluster as observed by WFXT in the deep
  (400 ksec) survey, at redshifts 0.5 (left), 1.0 (middle), 1.5
  (right). Images sizes are 10x10 arcmin.}
\label{bullet}
\end{figure*}
In this regime (i.e. HEW=5\arcsec), the degradation of the $c_{SB}$
measurement due to the angular resolution is moderate and can be
accounted for, while for angular resolution approaching 10\arcsec,
this effect rapidly increases and make it impossible to measure the
cool-core strength (see Fig.~\ref{wfxt2}).

Besides detecting and characterizing distant cool cores, WFXT's
angular resolution will also allow sharp features (such as cold front and
shocks) to be detected at high-z. This is illustrated with simulations
of the well-known Bullet cluster as it would appear in the WFXT deep
survey, at redshifts 0.5, 1.0 and 1.5 (see Figure 5).

\section{Conclusions}

In this contribution we investigated the evolution of cool-core
clusters across the entire redshift range currently available, i.e.,
out to $z$=1.3.  Our analysis is based on the archival
\textit{Chandra} data of three cluster samples, and our results are
derived mainly from the cluster X-ray surface brightness properties.
The distributions of the surface brightness concentration $c_{SB}$
(Fig.~\ref{csb}) show us that: (i) the 400SD and the RDCS+WARPS high-z
samples are statistically different: the 400 SD high-$z$ sample
appears to miss concentrated clusters; (ii) the distribution of
cool-core strength in the local and the RDCS+WARPS samples is rather
similar, even though the distant sample lacks very peaked (or strong
cool-core with $c_{SB}$ $>$0.15) clusters.

The distribution of the central cooling time in the
local sample spans a broad range, 0.7$<t_{cool}<$32.6 Gyr, where
two-thirds of the sample have $t_{cool}<$ $t_{Hubble}$.  The
RDCS+WARPS sample shows a somewhat different behaviour, displaying a
narrower range of cooling times, ([4.7-14.3] Gyr), and a median
$t_{cool}\sim$ of 8.9 Gyr. We confirm a strong anti-correlation
between $c_{SB}$ and $t_{cool}$, quantified by a Spearman rank
coefficient of $\rho$=$-$0.84.

Our results extend the current knowledge of the cool-core population
to the most distant X-ray clusters known to date, and show that even
at such large lookback times, we detect a significant population of
well developed cool cores.  A significant advancement in this research
can only be achieved when large samples will be available. This will
be possible only with the next generation X-ray survey missions.  In
particular, we showed that WFXT will have the capacity to resolve the
central regions of strong cool-cores up to redshifts $z\sim 1.5$.
Since WFXT is expected to yield hundreds of new cluster detections at
$z \sim 1$, it will add significant constraints to the formation and
evolution of cool-cores in galaxy clusters.

\bibliographystyle{aa}

\end{document}